\documentclass[]{spie}  

\usepackage{amsmath,amsfonts,amssymb}
\usepackage{graphicx}
\usepackage[colorlinks=true, allcolors=blue]{hyperref}

\title{Towards precision particle background estimation for future X-ray missions: correlated variability between \\ Chandra ACIS and AMS}

\author[a]{Catherine E. Grant}
\author[a]{Eric D. Miller}
\author[a]{Marshall W. Bautz}
\author[a]{Richard Foster}
\author[b]{Ralph P. Kraft}
\author[c]{Steven Allen}
\author[d]{David N. Burrows}

\affil[a]{MIT Kavli Institute for Astrophysics \& Space Research, Cambridge, Massachusetts, USA}
\affil[b]{Center for Astrophysics $|$ Harvard \& Smithsonian, Cambridge, Massachusetts, USA}
\affil[c]{Department of Physics, Stanford University, Stanford, California, USA}
\affil[d]{The Pennsylvania State University, University Park, PA, USA}

\authorinfo{Further author information: (Send correspondence to C.E.G.)\\C.E.G.: E-mail: cgrant@mit.edu}

\begin{document} 
\maketitle

\begin{abstract}

A science goal of many future X-ray observatories is mapping the cosmic web through deep exposures of faint diffuse sources. Such observations require low background and the best possible knowledge of the remaining unrejected background. The dominant contribution to the background above 1-2 keV is from Galactic Cosmic Ray protons. Their flux and spectrum are modulated by the solar cycle but also by solar activity on shorter timescales. Understanding this variability may prove crucial to reducing background uncertainty for ESA's Athena X-ray Observatory and other missions with large collecting area. We examine of the variability of the particle background as measured by ACIS on the Chandra X-ray Observatory and compare that variability to that measured by the Alpha Magnetic Spectrometer (AMS), a precision particle detector on the ISS. We show that cosmic ray proton variability measured by AMS is well matched to the ACIS background and can be used to estimate proton energies responsible for the background. We discuss how this can inform future missions.
\end{abstract}


\section{INTRODUCTION}
\label{sec:intro}  

Mapping faint diffuse structures such as clusters and groups of galaxies, and the intergalactic medium is a science goal of many future X-ray observatories and of the “Cosmic Ecosystems” theme of the Astro2020 Decadal Report ``Pathways to Discovery in Astronomy and Astrophysics for the 2020s"\cite{Astro2020}. Such observations require low background and the best possible knowledge of the remaining unrejected background. For most detectors and orbits, the dominant contribution to the unrejected background above 1-2~keV is from Galactic Cosmic Ray (GCR) protons. The flux and spectrum of these protons are modulated by the 11-year solar cycle but also with solar activity on much shorter timescales of months, weeks, or even days. Understanding this variability may prove crucial to reducing background uncertainty to the level required by ESA's Athena X-ray Observatory\cite{athena} and other proposed future large collecting area missions like NASA's Lynx\cite{lynx} and the probe concept Advanced X-ray Imaging Satellite (AXIS)\cite{AXIS}.

These future missions are designed with ambitious goals in mind---finding the first groups and clusters of galaxies, and mapping cosmic structures in time---to take deep exposures and measure the properties of faint diffuse objects.  Any source of background, whether due to particles, stray light, unresolved sources, or optical light leak, adds statistical and systematic uncertainty to the final measurements of temperature, density, etc. Observations of very low surface brightness sources are often currently limited by systematic uncertainty in the background level. Measuring the background in an ``empty" portion of the observation field of view is not always possible for extended emission sources, so the background must be estimated by scaling from other observations separated in time. That scaling factor adds additional systematic error to the final results. Meeting the ambitious goals of future missions will require both reducing the background as much as feasible and improving our knowledge of the remaining background.

\begin{figure}[t]
\begin{center}
\includegraphics[width=4.0in]{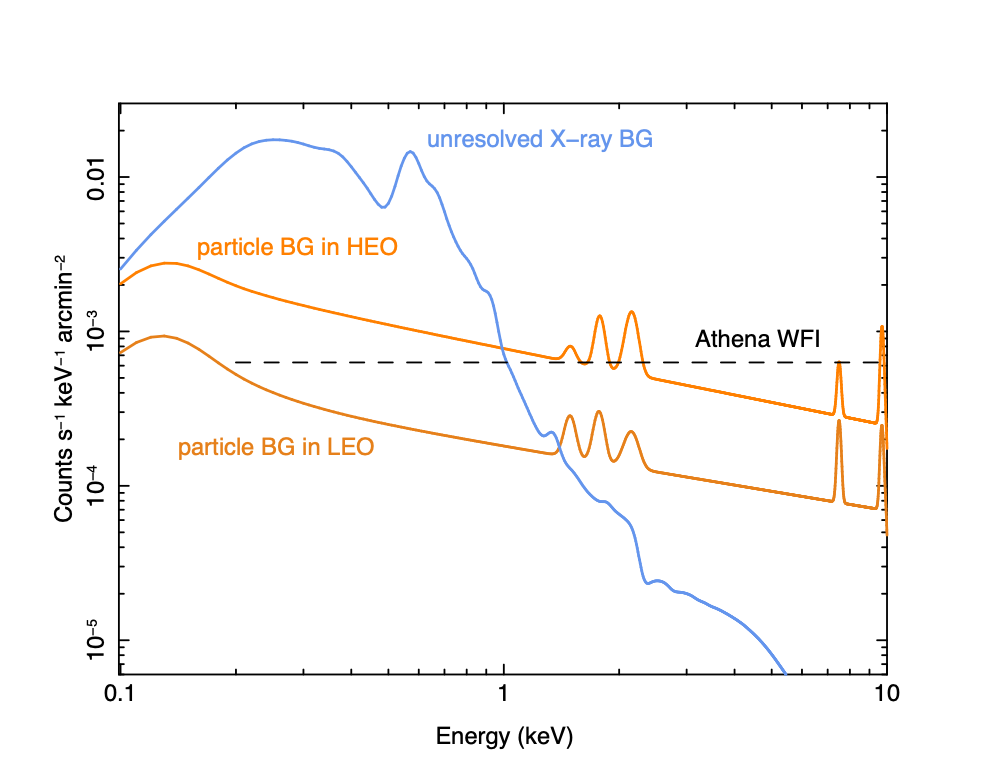}
\caption{Simulated spectra comparing the unresolved X-ray background with the particle background in high-Earth orbit (HEO) and low-Earth orbit (LEO) for one proposed observatory configuration. In this example, the particle background is important above 1 to 2~keV. The normalization of the background is different in LEO and HEO, due to the geomagnetic shielding present in LEO. The dashed line is the current background requirement for the Athena WFI.}
\label{fig:xrbfig}
\end{center}
\end{figure}

There are many components of the background seen in X-ray observations. One way to subdivide them is into two groups: focused X-rays, i.e. unresolved point sources and the diffuse X-ray background, and everything else. For most detectors, observatories, and orbits, the focused X-rays dominate the background below 1 to 2~keV (Figure~\ref{fig:xrbfig}). The particle background itself has a number of sources: Galactic cosmic rays (GCRs), solar energetic protons, and the Earth's trapped radiation.  This part of the background is temporally variable and its normalization depends on the presence or absence of geomagnetic shielding. In addition, there's an unfocused cosmic hard X-ray background which produces a constant pedestal in the measured non-focused background. Outside of solar storms and the Earth's radiation belts, the primary contribution to the non-X-ray background is GCR protons, either by direct interaction in the detector, or by secondaries produced in interactions within and around the detector.

The choice of orbit for a proposed mission, while often primarily driven by constraints other than the particle background, has major ramifications for its overall normalization and the variability. In low-Earth orbit (LEO), the GCR background is modulated by changing geomagnetic shielding along the orbit. An example is Ref.~\citenum{xisbkg} in which the cutoff rigidity along the orbit can be used to model the non-X-ray background of the XIS on Suzaku with good accuracy. In high-Earth orbit (HEO), most or all of the science observing time occurs while the spacecraft is well outside radiation belts and there is no geomagnetic shielding. Both NASA's Chandra X-ray Observatory\cite{cha2,acis} and ESA's XMM-Newton\cite{xmmmos,xmmpn} have more than two decades of experience with the particle background in HEO. The normalization of the background is higher than in LEO and the variability is primarily driven by the solar cycle and solar activity. Neither of these can be simply or accurately predicted in the same way that orbit and Earth's magnetic shielding can be used for prediction in LEO. Knowledge of the background level and variability needs to come from other measurements.

(There is an additional time-varying component of the non-X-ray background as seen by XMM-Newton and, to a lesser extent, Chandra that has thus far been ignored: soft proton flares, which are brief enhancements of focused protons with energies of order 100~keV\cite{deluca2004}. This component is still not well understood but is believed to be associated with the Earth's magnetopause\cite{kuntz2008} and should therefore be less important in L1/L2.  As we are primarily interested in GCR protons and are utilizing stowed ACIS data, we will not be addressing soft protons.)

The launch of SRG/eROSITA in 2019 begins a new era as the first X-ray mission to orbit the Sun-Earth Lagrangian L2.\cite{erosita} The expectation was that the particle background in L2 should look much the same as that seen in HEO by XMM and Chandra, with the possible exception of soft proton flares.  While there are some questions about the normalization, which is higher than was expected based on pre-launch Geant4 modeling, for the most part this appears to be true. Ref.~\citenum{erositabkg} reports the background has been declining as solar activity increased, and found a $\sim27$ day period in the particle rate, possibly related to solar rotation.

Athena (Advanced Telescope for High ENergy Astrophysics), ESA's next large X-ray observatory, is scheduled to launch in the mid 2030s to Earth-Sun L1\cite{athena}. As some of the Athena science goals require study of faint diffuse emission, such as that in clusters of galaxies, there has been a substantial effort to better understand, model, and reduce the instrumental background produced by particles and high energy photons. The background reduction studies include detailed Geant4 simulations, experimental testing of shielding and coating choices, as well as software mitigation strategies, both on-board and on the ground.  A summary of these ongoing efforts by the Athena WFI team and the WFI Background Working Group is given in Ref.~\citenum{wfibkg}.  The work described in this proceedings paper was done as part of the US contribution to the Athena WFI and as part of the WFI Background Working Group.
  
We examine the variability of the particle background as measured by the Advanced CCD Imaging Spectrometer (ACIS) on the Chandra X-ray Observatory, and then compare that variability to measurements by the Alpha Magnetic Spectrometer (AMS), a precision particle physics detector on the International Space Station.  We show that the cosmic ray proton variability measured by AMS is very well matched to the ACIS variability and can be used to estimate the proton energies responsible for the ACIS unrejected background. We will discuss how this knowledge can inform future missions, by inclusion of special purpose particle monitors, as is under investigation for Athena, or by comparison to other space-based particle monitors.

Using correlated variability to learn more about the sources of the NXB has recently been done for both the XMM EPIC-pn\cite{xmmpnbkg} and EPIC-MOS\cite{xmmmosbkg} cameras, with comparisons to the on-board XMM Radiation Monitor, the EPHIN particle instrument on SOHO, and Chandra ACIS. After excluding solar energetic particle events, they find tight correlations between $\sim$1~GeV protons and all the X-ray background measures. The addition of the AMS to such comparisons with its large collecting area, higher spectral resolution, and high statistical precision, may allow for stronger conclusions on the proton energies of importance, and on the effectiveness of instrument proxies on background knowledge.

We begin in Section~\ref{sect:data} with a description of the ACIS instrumental proxy for the particle background, and of the AMS data. A comparison between the two is made in Section~\ref{sect:comp} and we conclude with a discussion of future work in Section~\ref{sect:disc}.

\section{DATA}
\label{sect:data}

\subsection{Chandra ACIS}
\label{sect:acis}

For this project we use data from the Advanced CCD Imaging Spectrometer (ACIS)\cite{acis} on the Chandra X-ray Observatory\cite{cha2}, which was launched into a highly elliptical 2.6-day orbit in 1999. 
Chandra's orbit samples a variety of particle environments, but spends most of its time above the Earth's radiation belts. It is expected that the unfocused particle background experienced by any future mission in HEO or L1/L2 should be very similar to that seen by Chandra. GCR protons are the primary source of quiescent background, modulated by the solar cycle and solar activity.  The twenty-three year baseline provided by Chandra data covers more than half of solar cycle 23, all of solar cycle 24, and the beginning of solar cycle 25, allowing for study of conditions at solar maximum and minimum and for variances between solar cycles.

ACIS utilizes framestore charge-coupled devices (CCDs), ten CCDs in total, each with 1024 by 1024 24-micron pixels. We focus on the back-illuminated (BI) CCD, ACIS-S3, at the ACIS-S aimpoint, as the closest analog to future detectors like the Athena WFI.  It is 45-microns thick, fully depleted, and runs in photon-counting mode, with frametimes of $\sim$3~second. Telemetry limitations require discarding the contents of most pixels on-board.  On-board processing identifies potential X-ray event candidates and telemeters the subset most likely to be X-rays and not particles. On-board filtering removes events with pulseheights above an upper threshold and event morphologies that including many contiguous pixels. While the filtered events themselves are not telemetered, the instrument does keep an accounting of the numbers of events removed and why. We use one of these counters, the high energy reject rate, as a proxy for the particle rate seen by the spacecraft.

\begin{figure}[t!]
\begin{center}
\includegraphics[width=6.5in]{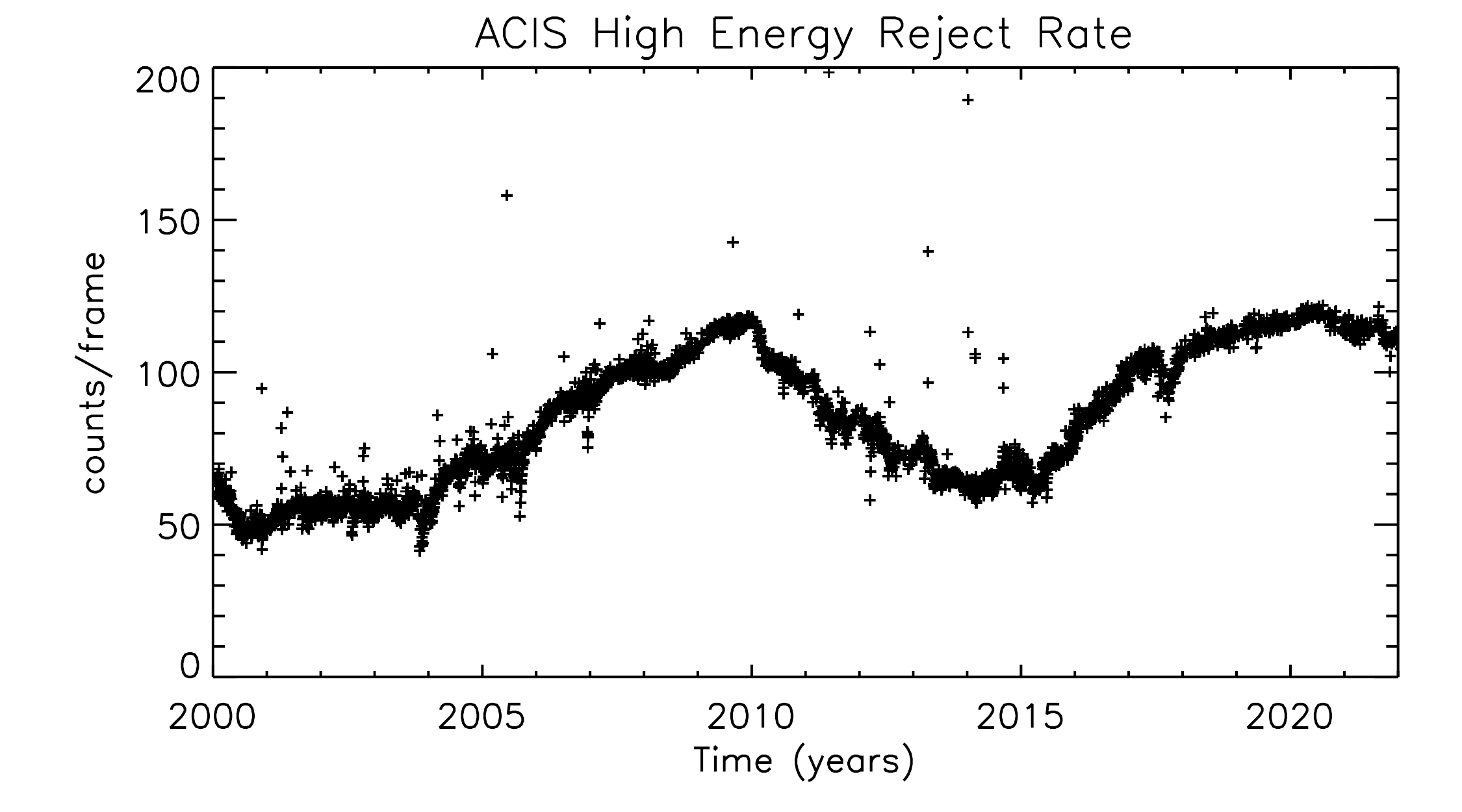}
\includegraphics[width=3in]{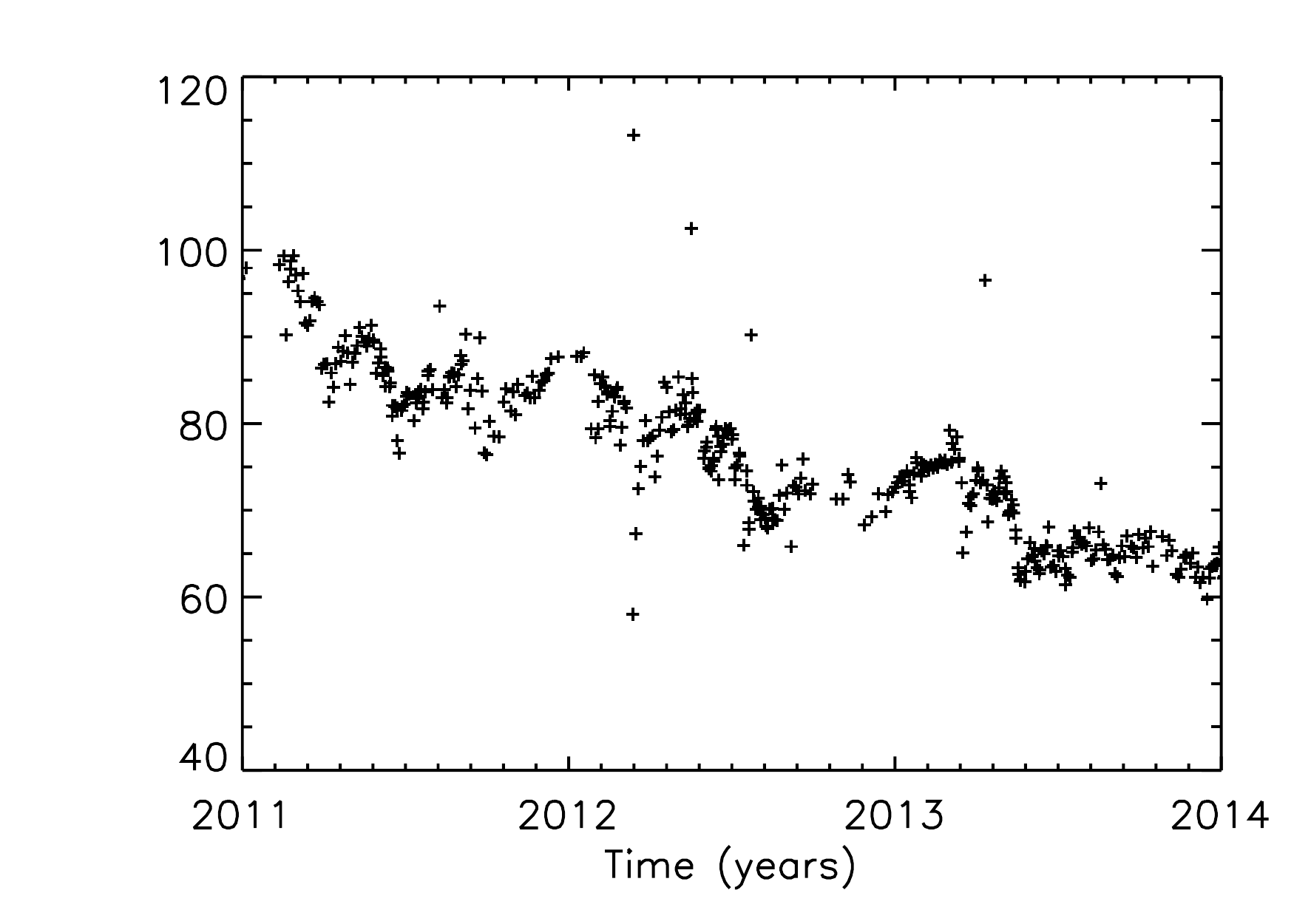}
\includegraphics[width=3in]{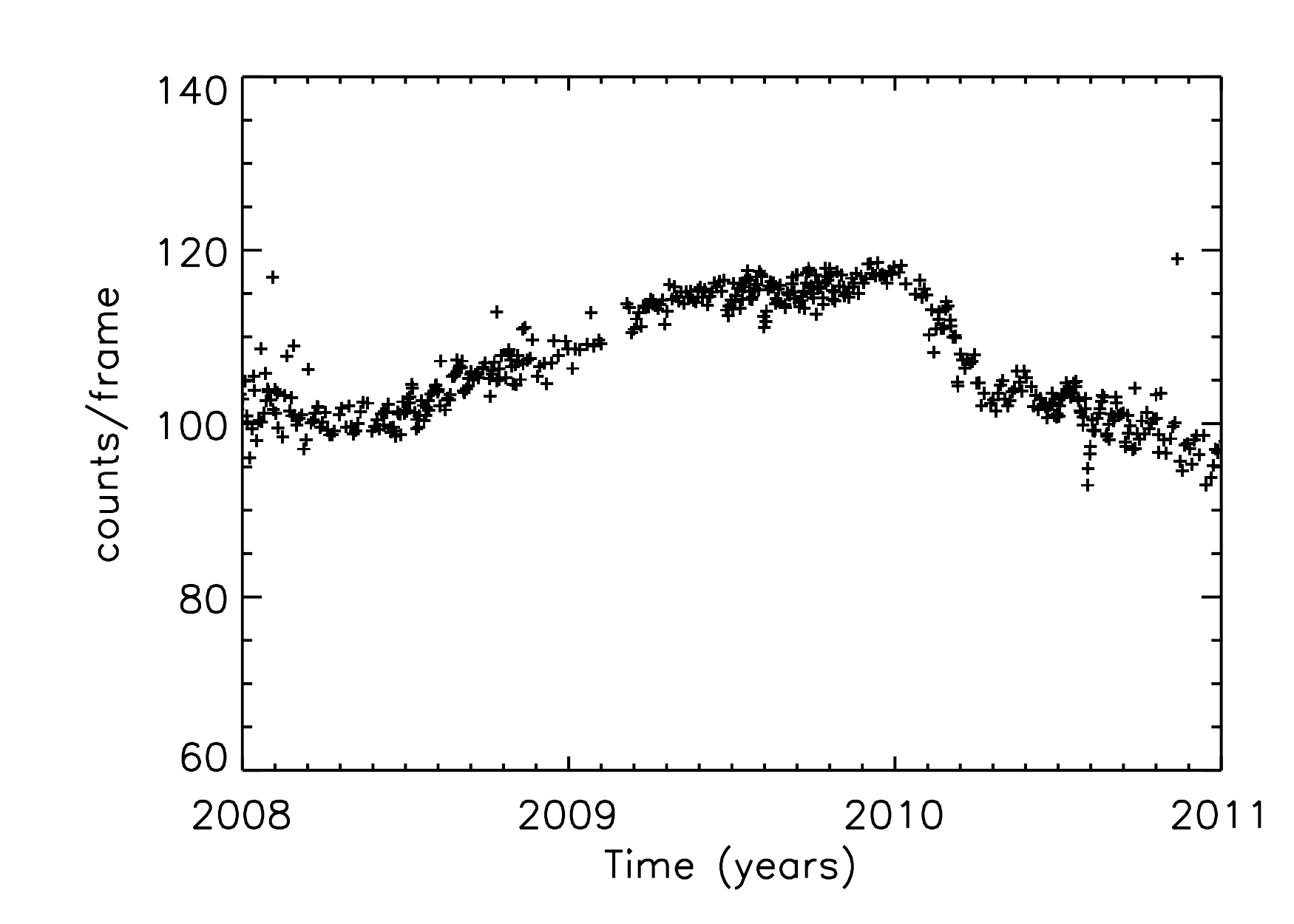}
\caption{Time dependence of the particle background as measured by the rate of high energy events on ACIS-S3. Each data point is a single ECS observation. (top) The full mission history of high energy rates on S3 includes two solar maxima and minima. The particle background is anti-correlated with the solar cycle and varies by a factor of $\sim$2 over 11~years. The two cycles are distinct; the first could not be used to accurately predict the rates in the next cycle. (bottom left) A segment of the ACIS-S3 background during solar maximum.  The sawtooth pattern is typical of solar activity and individual solar storms. The background variation is $\sim$10\% over weeks to months. (bottom right) A segment of the ACIS-S3 background during solar minimum, when there was no significant solar activity.  Background variations are smaller, of order 5\% over a matter of days.}
\label{fig:acisbkg}
\end{center}
\end{figure}

The data used here were taken while ACIS was stowed, viewing its External Calibration Source (ECS).  No focused X-rays reach the detector---only particles and fluorescent X-ray lines from the ECS, which are all well below the high energy reject limit, set to $\sim$15~keV during these observations. While the instrument is stowed, the background is dominated by Galactic cosmic rays, primarily protons $>$~10~MeV, plus any secondary particles produced through interaction with the spacecraft and instrument.  Focused low energy protons ($\sim$100~keV) which can cause short term increases in background count rates are not seen while the instrument is stowed.  These data are regularly taken just before entry and just after exit from the radiation belts, and are occasionally replaced by specialized instrument tests or are dropped entirely to improve science observing efficiency.  The observations are therefore not evenly spaced in time and the coverage is sometimes irregular.

The full history of stowed ACIS-S3 high energy reject rate is shown in the top panel of Figure~\ref{fig:acisbkg}. Each data point is the average rate for a single 3-10~ksec observation: the mean (and median) exposure time is 8~ksec. The statistical error on each point is smaller than the symbol size. The high-energy background measured by ACIS is clearly anti-correlated with the solar cycle and varies by a factor of $\sim$2 over 11~years. The solar maxima in 2001 and 2014 correspond to minima in background rates and vice versa. The solar cycles seen by ACIS are distinct; the first could not be used to accurately predict the rates in the next cycle.

The lower panels in Figure~\ref{fig:acisbkg} show shorter time sequences during solar maximum and solar minimum.  During solar maximum, a sawtooth pattern is caused by solar activity and individual solar storms. The background variation is $\sim$10\% over timescales of weeks to months. During solar minimum the background variations are smaller, but small scale ``bubbling" continues at the 5\% level over a matter of days.

All of this variation, large and small, long and short timescales, is relevant for modeling and removing the background contribution to scientific studies. Reducing the uncertainty in our knowledge of this variation would benefit any measurement of low surface brightness objects. Most detectors do a have a proxy of the background rate, much like the ACIS high energy reject rate, which does give an in situ simultaneous measure of the radiation environment, but generally the uncertainty on these counters is still large enough to influence the overall uncertainty of the final results.

\subsection{AMS}
\label{sect:ams}

X-ray detectors are optimized, in both hardware and software, to efficiently and accurately measure X-ray signal. They are not designed to capture all the signal from a particle interaction and cannot measure particle characteristics very well, if at all.  Specialized particle instruments can measure temporal and spectral characteristics of the particles and determine their species. This is why many X-ray observatories include a dedicated particle detector as part of their instrument complement, such as the Radiation Monitor on XMM-Newton.

\begin{figure}[t]
\begin{center}
\includegraphics[width=6.7in]{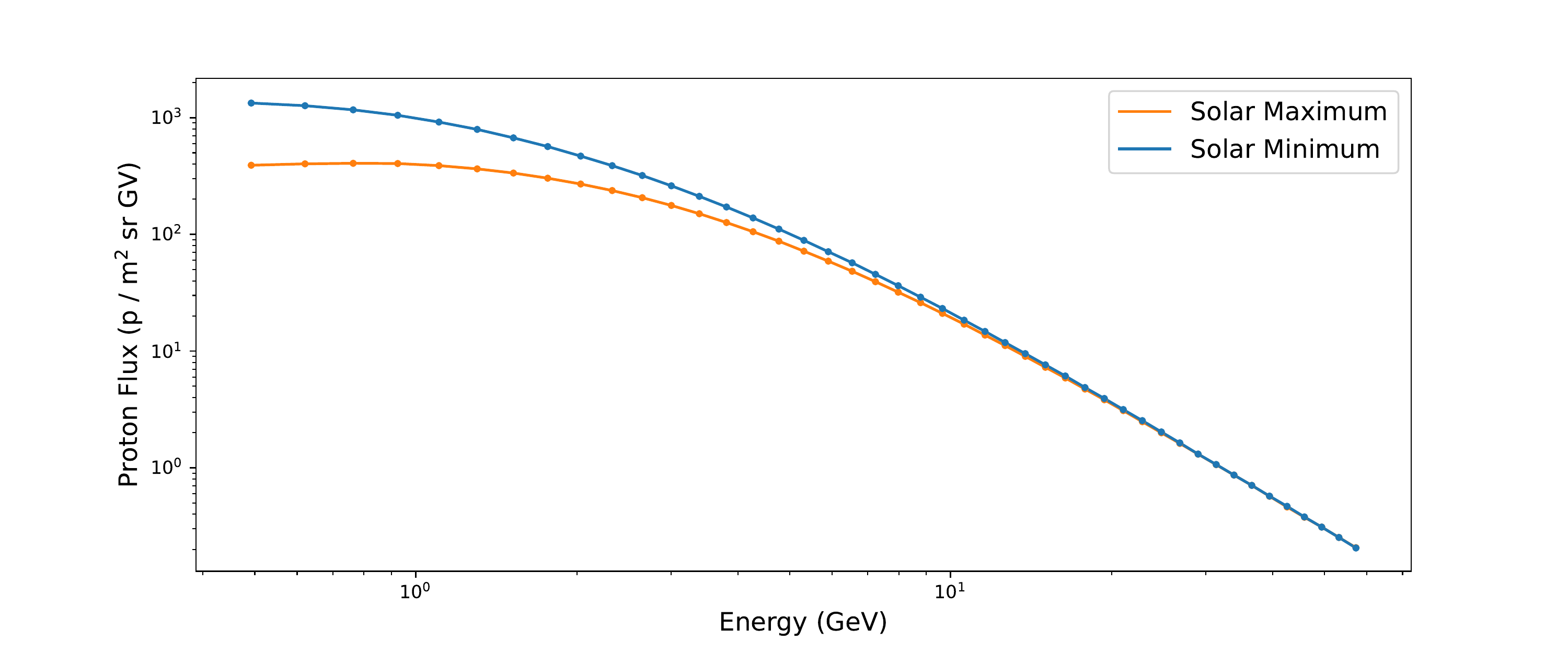}
\caption{AMS proton spectra at solar maximum (February 2014) and in the transition to solar minimum (February 2017). The overall flux is higher at solar minimum due to reduced magnetic shielding from the heliosphere. At solar maximum, the flux is suppressed, particularly at lower energies where the shielding is more effective. The statistical plus systematic errors on each data bin are too small to be seen on this plot.}
\label{fig:amsbkg}
\end{center}
\end{figure}

In this paper, we look at the Alpha Magnetic Spectrometer (AMS) on the International Space Station (ISS).
AMS is precision general purpose particle instrument, sensitive to a wide range of particle energies and types, well documented in a series of on-going papers and data releases.\footnote{See https://ams02.space for a full list of references and results} In particular, our interest is AMS measurements of Galactic cosmic ray protons, which are described in Ref.~\citenum{ams2015,ams2018,ams2021}. The AMS team has extensively calibrated the instrument, developed a highly detailed simulator, and carefully verified all steps of the data analysis process. 

At this point the question could be asked, how can AMS tell us about GCR protons when it orbits in LEO? As already stated, our knowledge of the ISS orbit and the geomagnetic shielding experienced at any point in the orbit allows for filtering of the proton data, to include only those events with energies well above the cut-off rigidity at that orbital position.  As the ISS orbit traverses a wide range of cut-off rigidities, the GCR proton spectrum, as it would be seen outside geomagnetic influence, can be reconstructed. The AMS team has done such a reconstruction, and includes estimates of the systematic error introduced by this and all the other calibration and filtering steps. AMS is well suited as an instrument for exploration of the particle background experienced by any X-ray instrument.

Each publication by the AMS team is accompanied by a public release of the relevant calibrated data.  For GCR protons, there have been three releases: first, the precise measurement of the proton spectrum;\cite{ams2015} next, the temporal behavior of proton fluxes as a function of Bartels rotation (27~days, or approximately the synodic solar rotation period);\cite{ams2018} and finally, the daily behavior of proton fluxes.\cite{ams2021} We have initially focused on the data binned by Bartels rotation\cite{ams2018}, and are beginning to explore the daily binned data.\cite{ams2021} The comparison to the daily binned data will be reported in a subsequent publication.

Figure~\ref{fig:amsbkg} shows an example of time-resolved spectra for two of the Bartels rotation bins. The first is in February 2014, during solar maximum, and the second is three years later, in February 2017, as rates are increasing towards solar minimum. The overall flux is higher at solar minimum due to reduced magnetic shielding from the heliosphere. At solar maximum, the flux is suppressed, particularly at lower energies where the shielding is more effective. The statistical plus systematic errors on each data bin are too small to be seen on the plot.  AMS proton data extend to much higher proton energies, of order 1~TeV, but those data are not shown here due to low fluxes and lack of any variability from heliospheric shielding.

\section{Comparison of ACIS to AMS}
\label{sect:comp}

\begin{figure}[t]
\begin{center}
\includegraphics[width=6.5in]{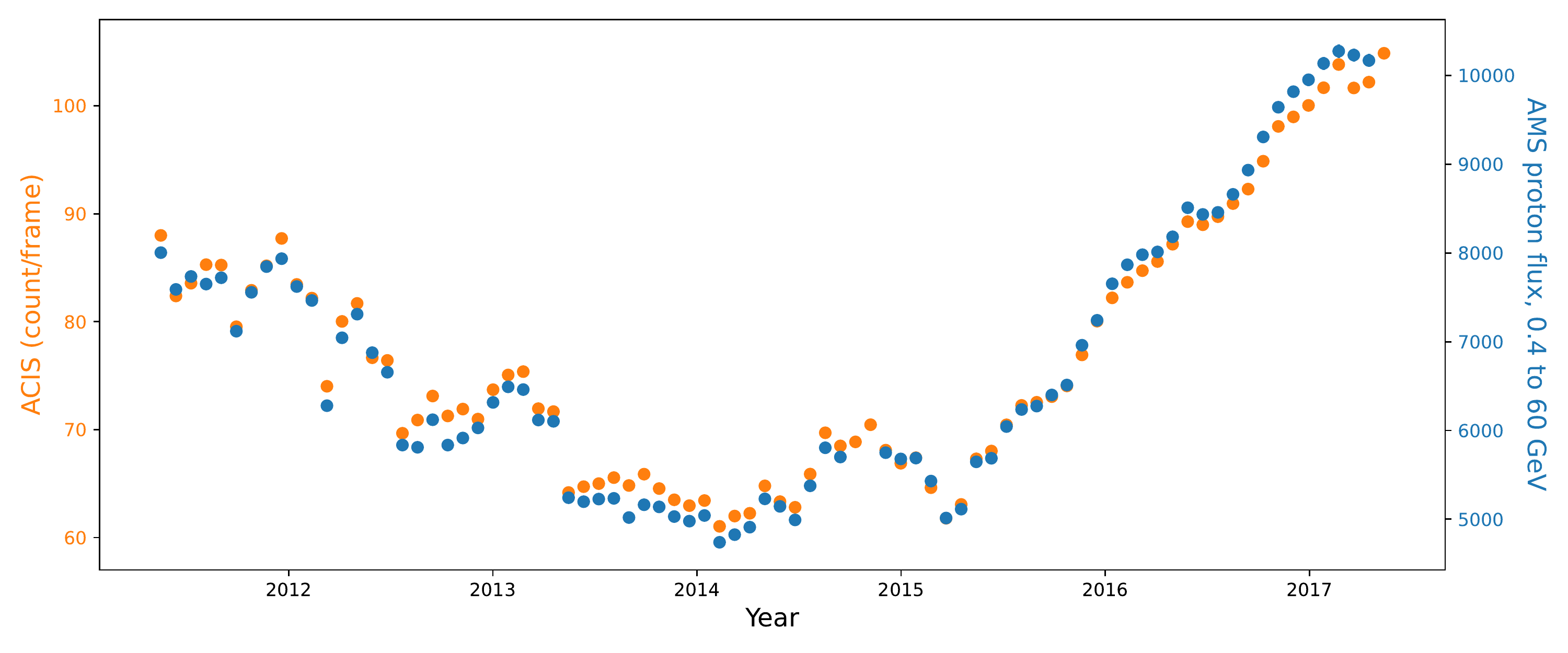}
\caption{A comparison of the ACIS-S3 high energy reject rate (orange) and the AMS proton flux (blue). The ACIS data have been rebinned to match the Bartels (27 day) binning of the AMS proton flux data. The 1-sigma error bars are plotted, but are too small to see. This plot does not represent a fit, just a scaling that allows both sets of data to fit in the same box. The correspondence between the two data sets is quite striking, both the overall shape and the shorter timescale features. }
\label{fig:acisams}
\end{center}
\end{figure}

Comparison of the ACIS-S3 high energy reject rate to AMS proton fluxes may help estimate the proton energies most responsible for the ACIS background and the strength of any correlation may indicate the effectiveness of the instrumental proxy on background knowledge. Before this comparison, however, the ACIS data requires filtering and rebinning. As seen in Figure~\ref{fig:acisbkg}, there is a locus of data points following the quiescent background level, but also numerous outliers above the quiescent rate. The outliers represent particular solar storms that elevated the particle rate.  As we are interested in the quiescent variations, these positive outliers were removed.  Negative outliers were not removed, since these represent longer term suppression of the particle background after particularly vigorous solar storms. Thirteen ACIS observations were removed, leaving a total of 1,044 observations over the time period covered by the AMS 27-day binned data.

Figure~\ref{fig:acisams} is a first comparison between the ACIS high energy reject rates and the AMS proton fluxes from 0.4 to 60 GeV, both binned to the 27-day Bartels rotation period. This plot does not represent a fit, just a scaling that allows both sets of data to fit in the same box. One-sigma error bars are included, but are too small to see (statistical plus systematic error for AMS, statistical only for ACIS). There is clearly a strong correlation, both in the long-term solar cycle changes and in the shorter timescale features. This plot also highlights how the shorter term behavior is different between the period with declining rates, with many sharp features, and the period with ascending rates, which is smoother. 

\begin{figure}[t]
\begin{center}
\includegraphics[width=6.5in]{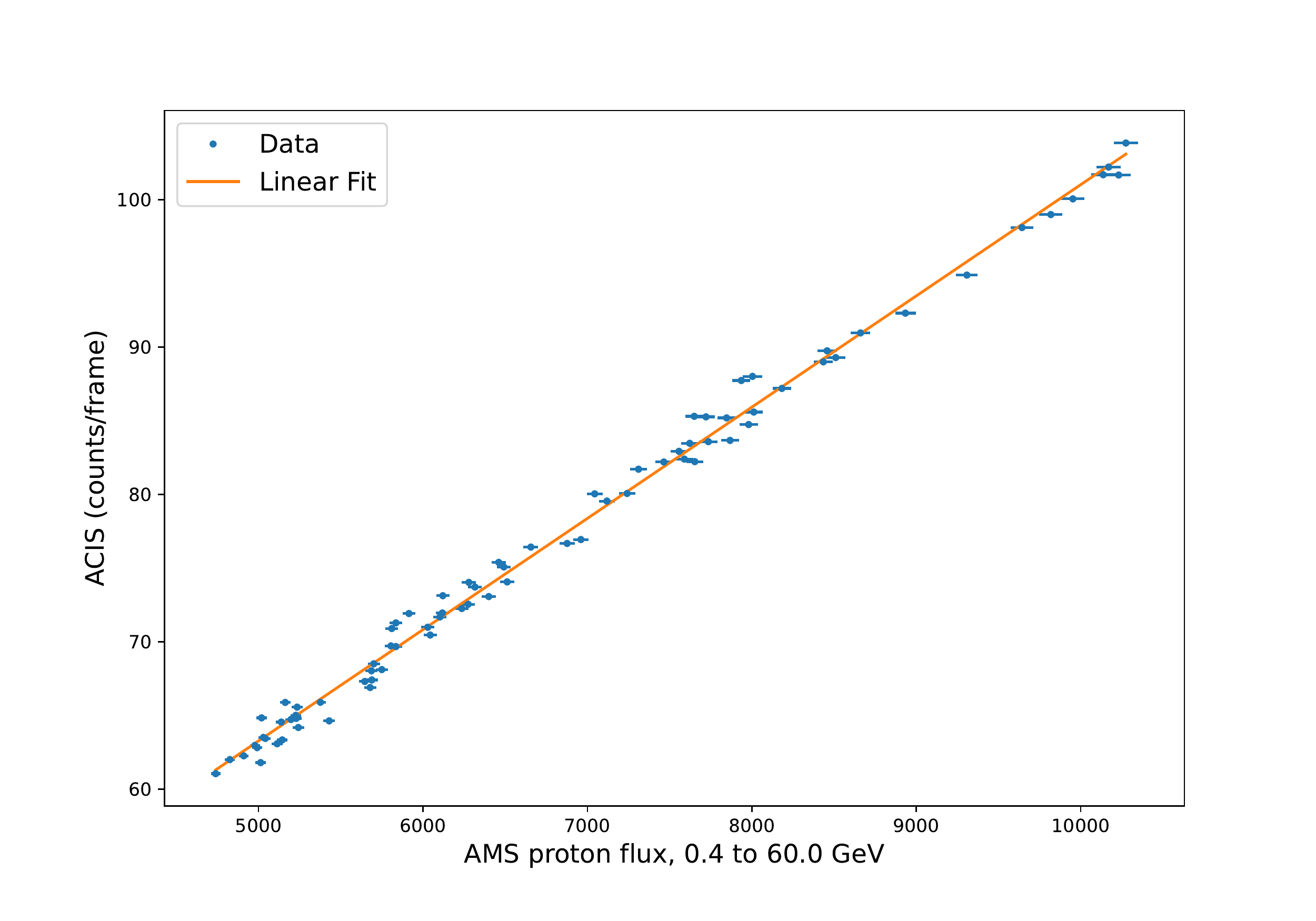}
\caption{ACIS-S3 high energy count rates as a function of AMS proton flux. The data are clearly well correlated. A linear fit is also shown. The scatter about the fit may be due to uneven sampling of the ACIS data within the time bin, or real differences between the two samples.}
\label{fig:acisams3}
\end{center}
\end{figure}

Figure~\ref{fig:acisams3} directly plots the ACIS high energy count rates against the AMS proton flux. This further demonstrates correlation between the ACIS background proxy and the actual particle measurements. A linear fit is also shown, which has a high degree of significance and no obvious curvature or other large scale deviation. There is some scatter from the fit, possibly due to the uneven sampling of the ACIS data within each time bin, or possibly real differences between the two samples. The linear fit, if extrapolated to zero proton flux, leaves a residual ACIS count rate of $\sim25$ counts/frame which is at least partially due to the unvarying pedestal provided by the hard X-ray background.

\begin{figure}[th]
\begin{center}
\includegraphics[width=6.5in]{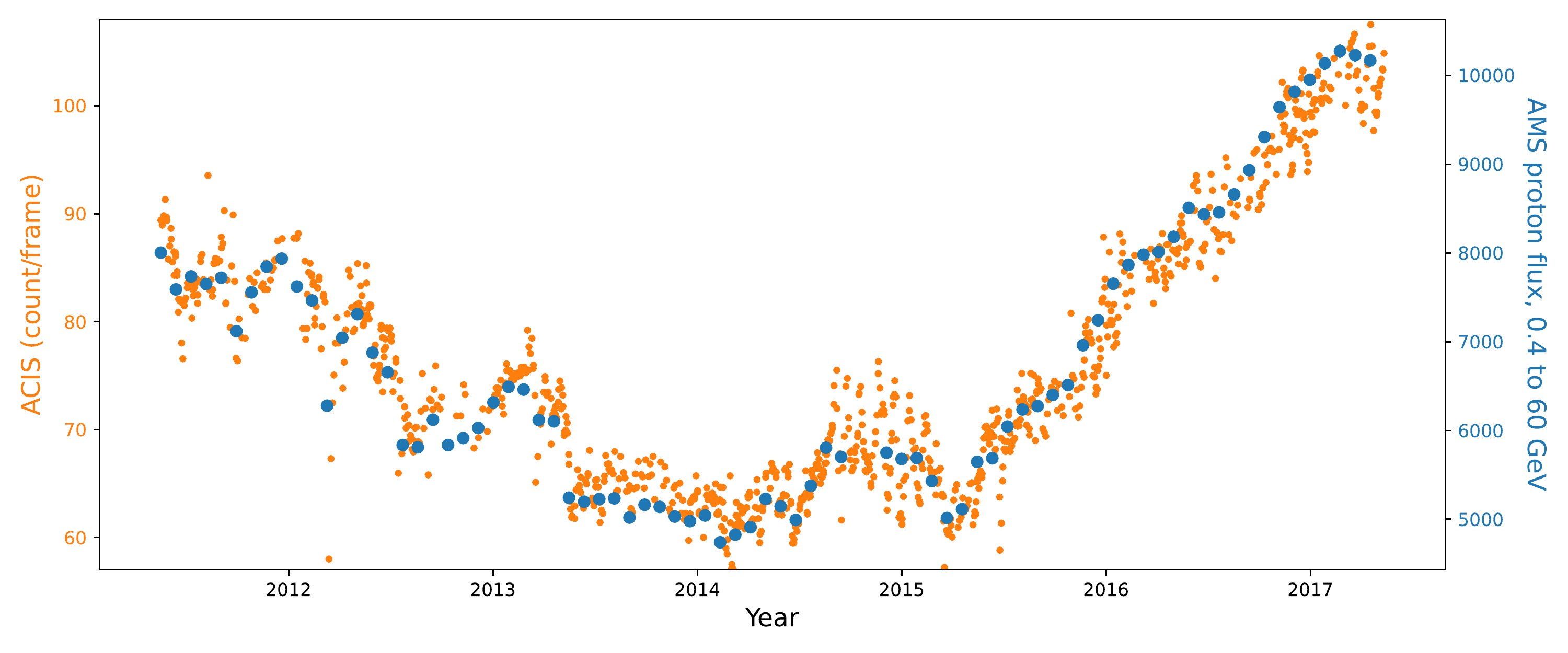}
\caption{A comparison of the ACIS-S3 high energy reject rate (orange) and the AMS proton flux (blue). The only difference between this figure and Figure~\ref{fig:acisams} is that the ACIS data are unbinned and show the size of the variation in the ACIS data within a 27 day Bartel rotation time bin.}
\label{fig:unacisams}
\end{center}
\end{figure}

To better understand how much of the scatter in the correlation is due to variation within each 27-day time bin, an additional comparison is shown in Figure~\ref{fig:unacisams}. This figure is quite similar to Figure~\ref{fig:acisams}, except that the original unbinned ACIS data are shown against the Bartels-rotation-binned AMS data.  While the correlation is still clear, there is substantial variation in the ACIS data on timescales shorter than 27~days. This is certainly an argument for examing the more recent AMS data with daily binning.

\begin{figure}[t]
\begin{center}
\includegraphics[width=6.5in]{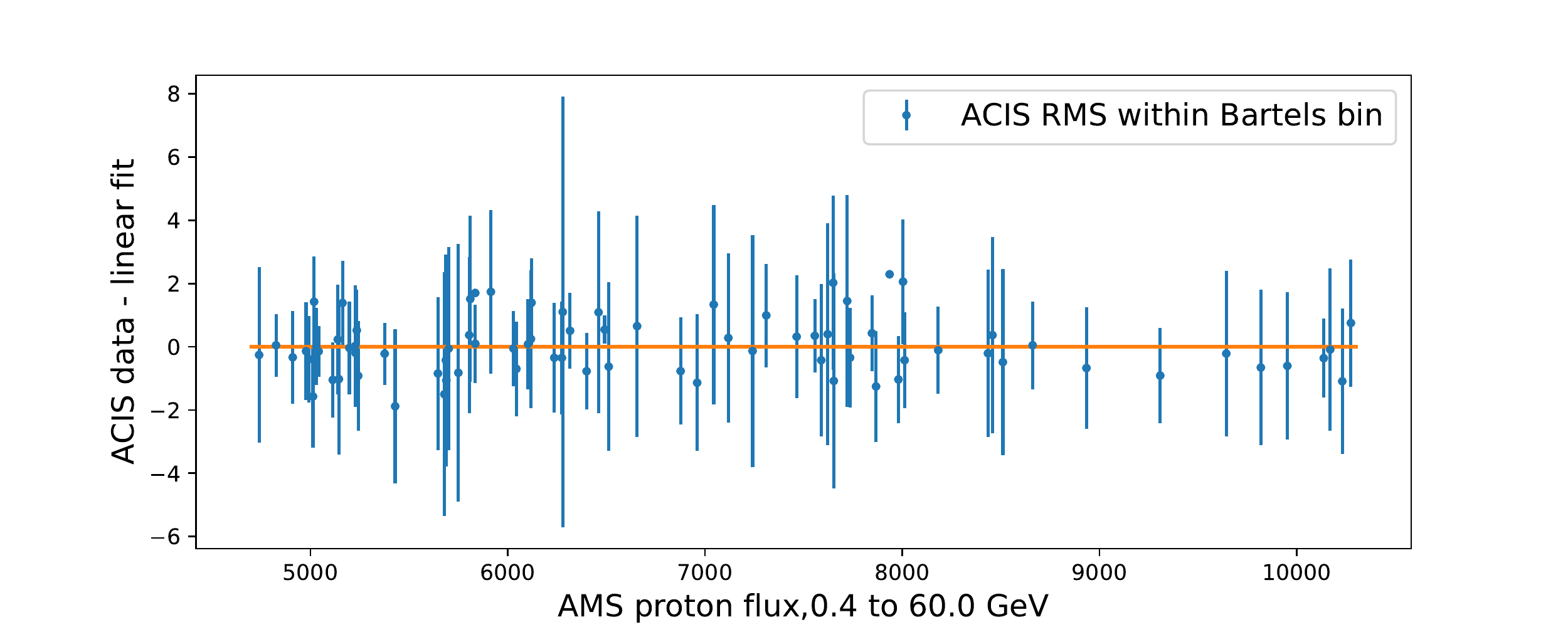}
\caption{The difference between the ACIS high energy count rate and the linear model, plotted as a function of the AMS proton flux. The data and the model are the same as that in Figure~\ref{fig:acisams3}, but instead of statistical error bars, the measured RMS of the individual observations in each bin is shown. The RMS within each bin is much larger than the scatter about the fit.}
\label{fig:acisams4}
\end{center}
\end{figure}

A further check on the scatter in the correlation is shown in Figure~\ref{fig:acisams4}.  Here the difference between the binned ACIS data and the linear fit is shown as a function of the AMS proton flux.  What's new is that instead of statistical error bars, this shows the RMS of the individual ACIS observations within each bin. The RMS is always larger than the difference between the ACIS data and the linear fit, indicating that the shorter timescale variations may explain some of the scatter in the correlation.

\begin{figure}[t]
\begin{center}
\includegraphics[width=6.5in]{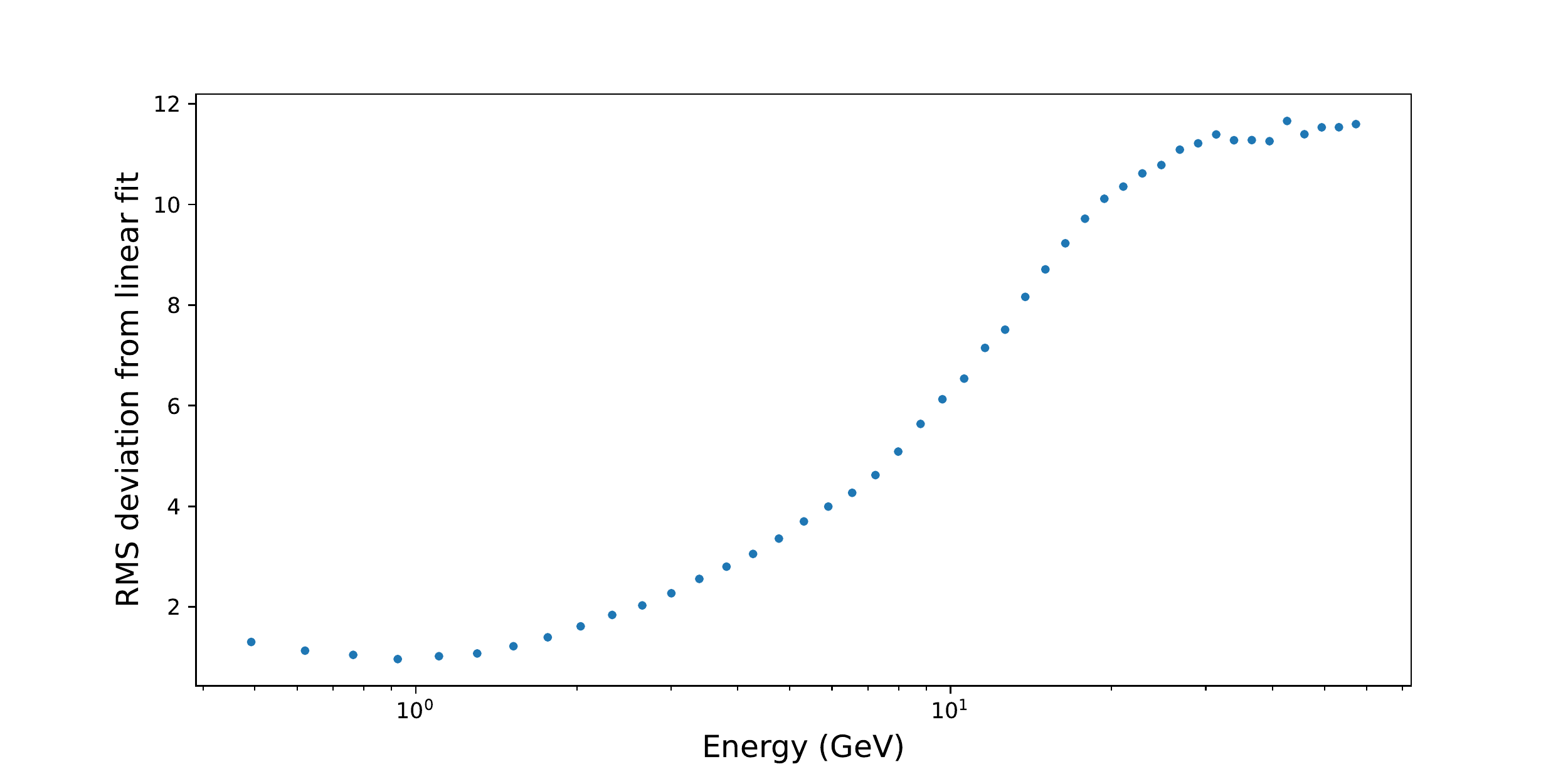}
\caption{RMS deviation from a linear fit to the ACIS high energy count rate and AMS proton fluxes, each proton energy bin separately. The best fits, lowest RMS values, are at the lower end of the energy scale around 1~GeV.}
\label{fig:acisams5}
\end{center}
\end{figure}

A final question that can be addressed is whether these data can be used to estimate the proton energies most responsible for the ACIS particle background. To address this, we perform linear fits to the binned ACIS high energy proxy and the AMS proton fluxes, but separately for each proton energy bin. If some proton energies are better correlated with ACIS, that can be an indication that those energies are more important in producing the ACIS particle background. In Figure~\ref{fig:acisams5} we show the RMS deviation from each of these linear fits, as a function of the proton energy in that bin. A minimum can be seen around 1~GeV, but there's a broader range of proton energies that can reasonably fit the ACIS data, from the lowest bin at 0.5~GeV to a few GeV. In reality, the ACIS focal plane experiences a range of proton energies, but protons around 1~GeV seem most important. This is in agreement with similar studies done with XMM\cite{xmmmosbkg,xmmpnbkg}.

\section{Conclusions and Future Work}
\label{sect:disc}

In this proceedings paper, we have compared a particle background proxy observed by ACIS on the Chandra X-ray Observatory with precision proton flux measurements by AMS on the ISS.  The ACIS background proxy shows a wide range of variability behaviors, from the long-term solar cycle to shorter periods of months or even days. The proton flux data also show variability, particularly for lower energies below 10~GeV or so, where the heliosphere magnetic shielding is important. For the proton data binned to the 27-day Bartels rotation period, we find a good correlation between the temporal behavior of the background proxy and the actual particle measurements. There is substantial variation of the ACIS background proxy data within each Bartel time bin, which may account for some of the scatter in the correlation. We also find that the fits are proton energy dependent, with a best fit around 1~GeV and acceptable fits from the lowest energy bin at 0.5~GeV to a few GeV.

This work is exploratory, intended to determine if further investigation is warranted. It is clear that AMS data are well suited for comparison to particle background measurements from X-ray detectors. An obvious avenue to pursue is comparison to the daily binned AMS proton fluxes, released late last year. This potentially may explain some of the scatter in the 27~day Bartels rotation binned data. It is also possible that some portion of the scatter is due real spatial differences in the particle environment seen by AMS and by Chandra. A question for future missions is whether in situ measurements by an onboard particle monitor are more valuable than higher precision measurements from a special purpose instrument like AMS. Further examination of the AMS data may help answer this and guide decision-making in the future.

\acknowledgments       
 We thank undergraduates Brian Avendano (MIT) and Alexander Yu (Oberlin) for the initial work on the Jupyter notebook used for this comparison. We are grateful to the AMS team for their thorough documentation and the public availability of AMS data. This work was done as part of the Athena WFI Background Working Group, a consortium including MPE, INAF/IASF-Milano, IAAT, Open University, MIT, SAO, and Stanford. We gratefully acknowledge support from NASA grant NNX17AB07G and NASA contracts NAS~8-37716 and NAS~8-38252.

\bibliography{report} 
\bibliographystyle{spiebib} 

\end{document}